\def\nn{\nu\bar{\nu}}
\def\bb{b\bar b}
\def\ttb{t\bar t}
\def\pp{p\bar p}
\def\qq{q\bar q}
\def\tl{l^\pm l'^\pm l^\mp}
\def\dl{l^+l^-\nn}
\def\lsdl{l^\pm l^\pm jj}

\def\ppH{\pp\rightarrow H}
\def\ppWH{\pp\rightarrow WH}
\def\ppZH{\pp\rightarrow ZH}

\def\met{\vec E\!\!\!\!/_T}
\def\ssb{$S/\sqrt{B}$}

\documentstyle[aps,epsf]{revtex}  

\begin{document}        

\baselineskip 14pt
\title{Higgs Prospects at the Upgraded Tevatron: Fermilab Study Results}
\author{John D. Hobbs \\
 for the Fermilab RunII SUSY/Higgs Working Group}
\address{SUNY -- Stony Brook}
%
\maketitle              

\begin{abstract}        
Preliminary results from a Fermilab study of the sensitivity for higgs
production at the Tevatron in run II and beyond are presented.  The study
extends existing results by systematically combining results for all decay
channels, considering the production of higher mass higgs bosons and
interpreting the results in the context of supersymmetric higgs production as
well as standard model production.  In addition new analysis methods which
significantly improve sensitivity are used.
\end{abstract}   	

\section{Introduction}               
The standard model of particle physics has been studied with very high
precision over the course of the past ten years, and no significant deviations
have been found.  Despite this, our understanding of the origin of electroweak
symmetry breaking is still incomplete.  This arises in large part because the
only remaining undetected standard model particle, the higgs boson, mediates
electroweak symmetry breaking in the standard model.  The highest available
center--of--mass energy for the years 2000 to 2004 will be at the Fermilab
Tevatron $\pp$ collider with $\sqrt{s}=2.0$~TeV.  It is natural to explore the
sensitivity to higgs production at the Tevatron. This paper contains
preliminary results from a year--long study conducted jointly by the Fermilab
theory group and the CDF and D\O\ experiments.\footnote{For more information, 
see {\tt http://fnth37.fnal.gov/susy.html}.}  The goal is to quantify the
higgs discovery potential at the Tevatron in the coming run II and possible
extensions.  Results are presented as the luminosities required to exclude
higgs at the 95\% confidence level, or to establish either 3$\sigma$ or
5$\sigma$ excesses over predicted backgrounds.

The starting points for this study are the higgs mass constraints expected from
LEP2\cite{higgs-limits} and previous Fermilab studies\cite{Tev2K}\cite{TeV33}.
This study extends the previous Fermilab results by (1) including additional
standard model decays in the mass regions previously explored, (2) testing the
sensitivity for higgs masses $M_H>135$~GeV, (3) systematically combining results
from all channels, (4) interpreting the results as supersymmetric(SUSY) higgs
production and (5) considering additional decay modes arising from SUSY models.
In addition, a detector simulation was developed which gives significantly more
realistic event reconstruction than some of the previous studies used.

This paper has six sections.  The first describes the production and decay of
standard model higgs bosons and the simulations used in this study.  The second
and third sections contain results for standard model higgs production in the
mass ranges $90\le M_H<135$~GeV and $135<M_H<200$~GeV respectively.  The fourth
section presents the combination of the results in sections two and three.  The
fifth section describes the extension of the results to SUSY production.  The
last section describes studies of additional SUSY--specific decays,
particularly final states having four $b$--quarks.

\section{Production, Decay, Event Generation and Detector Simulation}
\begin{figure}[ht]	
\centerline{\hbox{\epsfxsize=3.25in\epsfbox{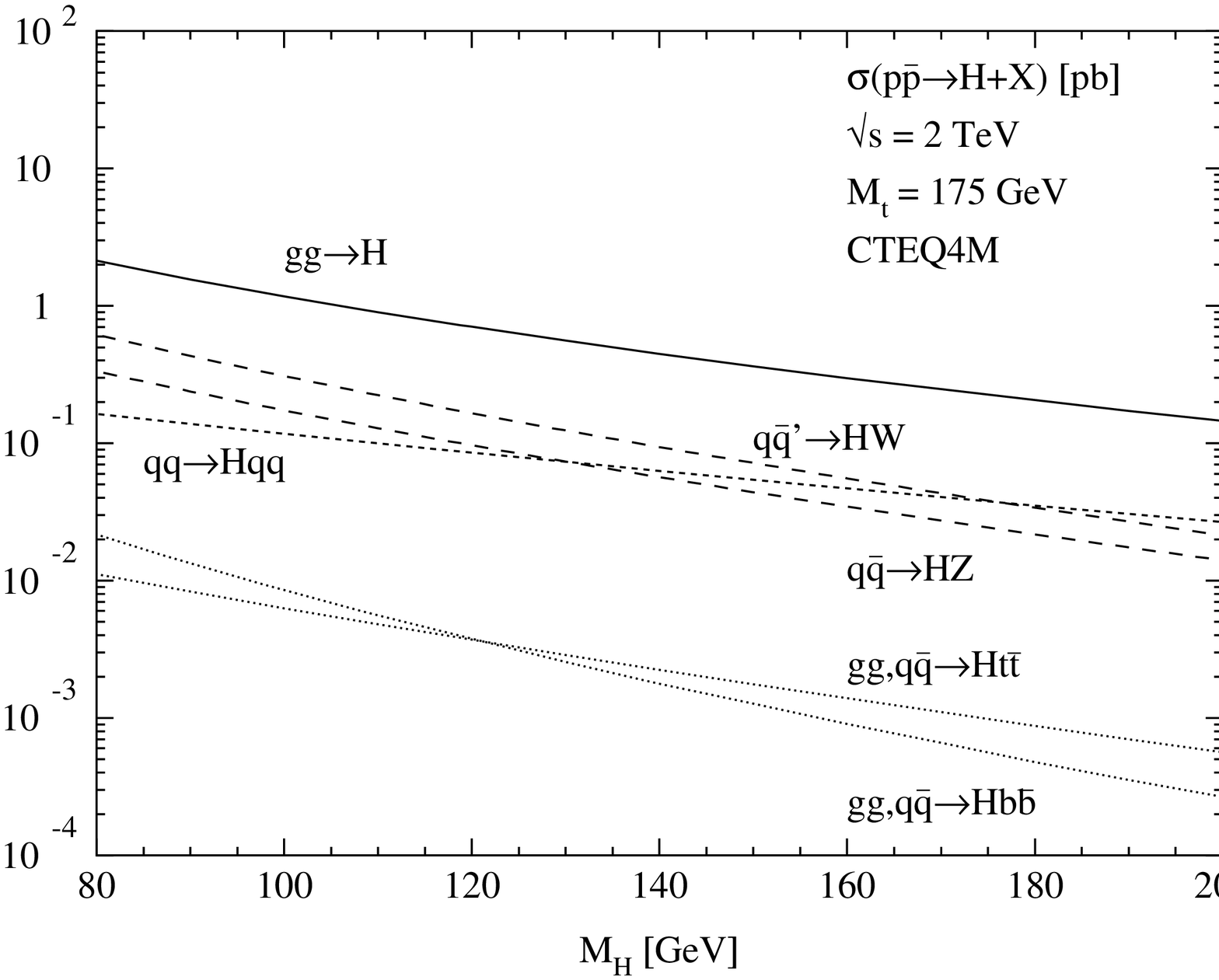}\hphantom{000000z}
   \epsfxsize=3in\epsfbox{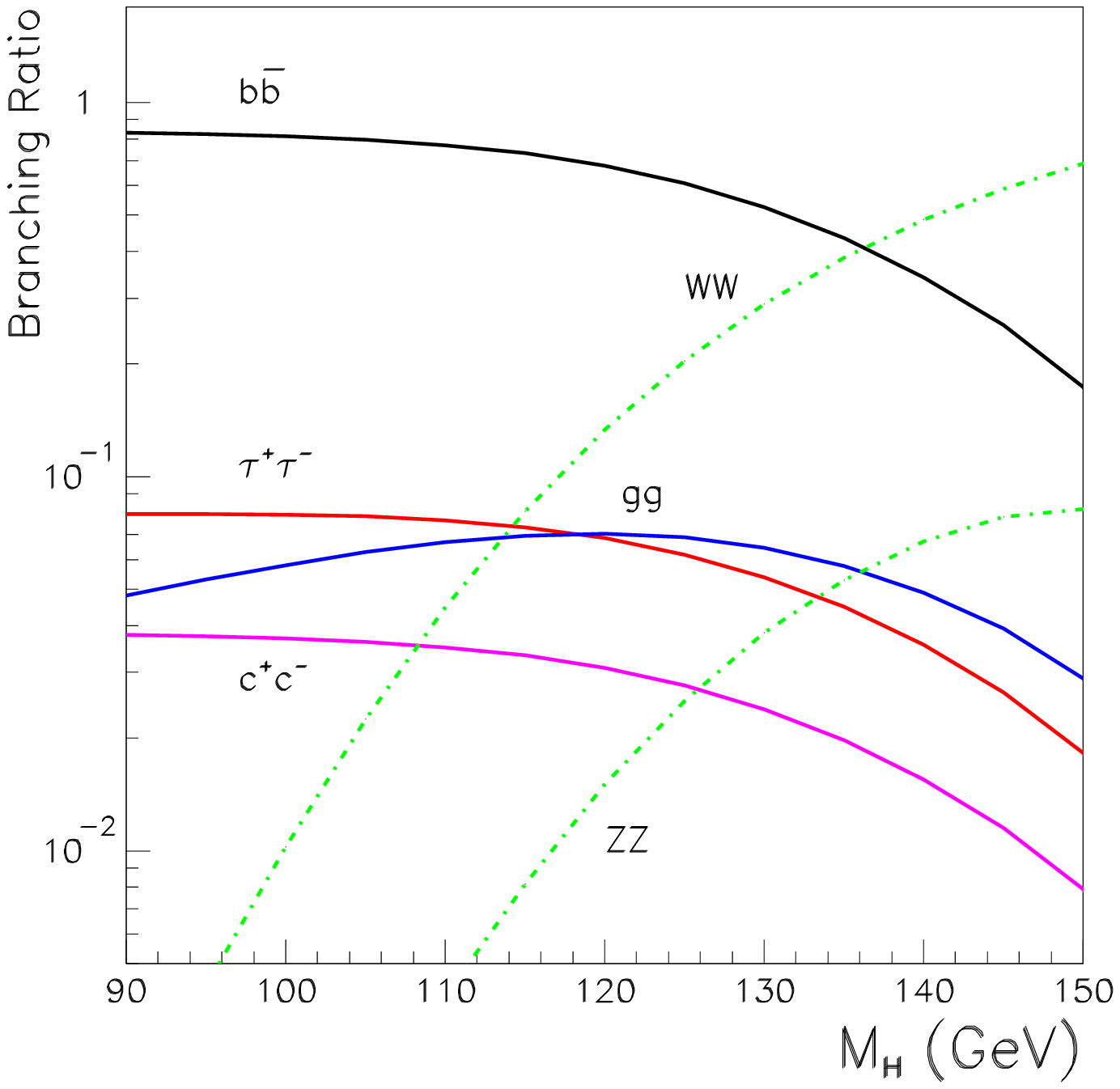}}}
\vskip -.2 cm
\caption[]{
\label{f-info}
\small Production cross sections and decay branching ratios for the standard 
  model higgs boson as a function of mass.}
\end{figure}

The production cross sections and decay branching ratios for a
higgs bosons have been calculated by a number of groups.\cite{spira}
  Those for a standard
model higgs boson are shown in Fig.~\ref{f-info}.  These plots indicate that
the highest cross section production modes are $\ppH$, $\ppWH$ and $\ppZH$.
The higgs decays dominantly to the most massive kinematically allowed final
state.  For $M_H<135$~GeV, the dominant decay mode is $H\rightarrow\bb$ with a
branching ratio of roughly 80\%.  For $M_H>135$~GeV, the dominant mode is
$H\rightarrow WW$.  Thus, searches for lower--mass higgs will be looking for
final states with at least two $b$--flavored jets, and the higher mass searches
will have multiple (virtual) $W$ bosons. For most of the mass range in
question, the $\ppH$ mode has very poor signal--to--noise, and the most useful
modes are the $\ppWH$ and $\ppZH$ modes with the $W$ or $Z$ decaying to
leptons.

Unless explicitly noted, events used in these analyses were generated using the
Pythia\cite{pythia}, Isajet\cite{isajet} or CompHep\cite{comphep} programs.
The generated four--vectors were then input to a detector simulation program,
SHW, developed for the run II workshop\cite{SHW}.  This program uses
parameterized resolutions for tracking and calorimeter systems and particle
identification to perform simple reconstruction of tracks, calorimeter--based
jets, vertices and trigger objects.  The resolutions used represent a typical
run II detector and are drawn from CDF and D\O\ internal studies.  Particle
identification efficiencies are included by parameterizing results from other
CDF and D\O\ studies.

The SHW program was verified by comparing selection efficiencies between SHW
and data or between SHW and well--established run I simulations used by CDF or
D\O.  The most stringent test was a comparison of nearly identical analyses of
the $\ppWH\rightarrow (\ell\nu)(\bb)$ channel.  Two analyses of this channel
have been performed, one based on a run I CDF simulation with the geometrical
acceptance extended to correspond to the run II CDF detector and the second
based purely on SHW.  The first analysis predicts $5.0$ signal events and
$62.8$ background events/fb$^{-1}$ for $M_H=110$~GeV.  The second predicts
$4.5$ signal and $62.5$ background events for the same conditions.

\section{Low Mass Higgs Searches, $M_H<135$~GeV}
When $M_H<135$~GeV, the dominant decay mode is $H\rightarrow\bb$.   Analyses
have been performed for all $\ppWH$ and $\ppZH$ final states.\footnote{The
mode $\ppH\rightarrow\bb$ was considered, but the signal to noise was too poor
for it to have any sensitivity when compared to the $\ppWH$ and $\ppZH$ modes.}
The possible final states are:
(1) $\ppWH\rightarrow\ell\nu\bb$,
(2) $\ppZH\rightarrow\nn\bb$,
(3) $\ppZH\rightarrow\ell^+\ell^-\bb$ and
(4) $\ppWH\rightarrow\qq\bb$ or $\ppZH\rightarrow\qq\bb$.
The primary backgrounds to these channels are $W+\bb$ and $Z+\bb$ with the
$\bb$ pair from gluon radiation, single top--quark production and top--quark
pair production.

All analyses for these channels begin with a preliminary selection based on
the number and type of final state objects.  For example, the 
$\ppWH\rightarrow\ell\nu\bb$ analysis requires a charged lepton with
$E_T>20$~GeV, missing transverse energy $\met>20$~GeV and two $b$--tagged
jets having $E_T>15$~GeV.   Similar selections are applied for the other
channels.  After the basic selection, a requirement is made that the mass of 
the reconstructed $\bb$ system be within (typically) 2$\sigma$ of the generated higgs mass.
Additional clean up requirements are also made.   As an example, in the $\ppZH
\rightarrow\nn\bb$ channel, there can be no isolated tracks with $p_T>15$~GeV.
This rejects events with high-$p_T$ leptons which failed the lepton
identification.  The resulting number of signal and background events
corresponding to 1~fb$^{-1}$ of data are given in table~\ref{t-lowmass}.  The
$\ppWH\rightarrow\ell\nu\bb$ and $\ppZH\rightarrow\nn\bb$ modes offer the best
sensitivity with the $\ppZH\rightarrow\ell^+\ell^-\bb$ mode not far behind.
The all--hadronic final state looks quite difficult.

\begin{figure}[ht]	
\hbox{\epsfxsize=3.25in\epsfbox{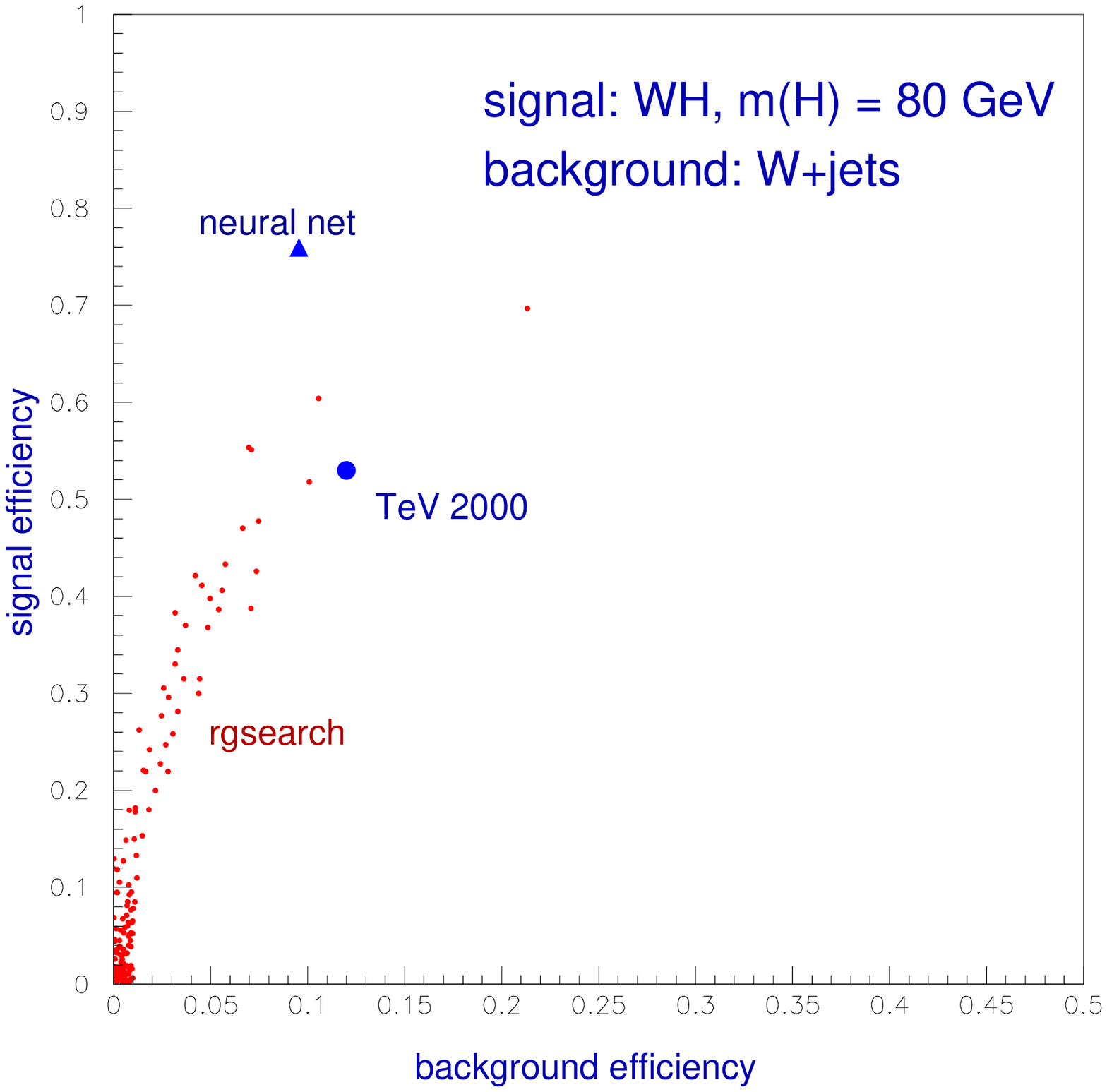}\hphantom{000}
   \epsfxsize=3in\epsfbox{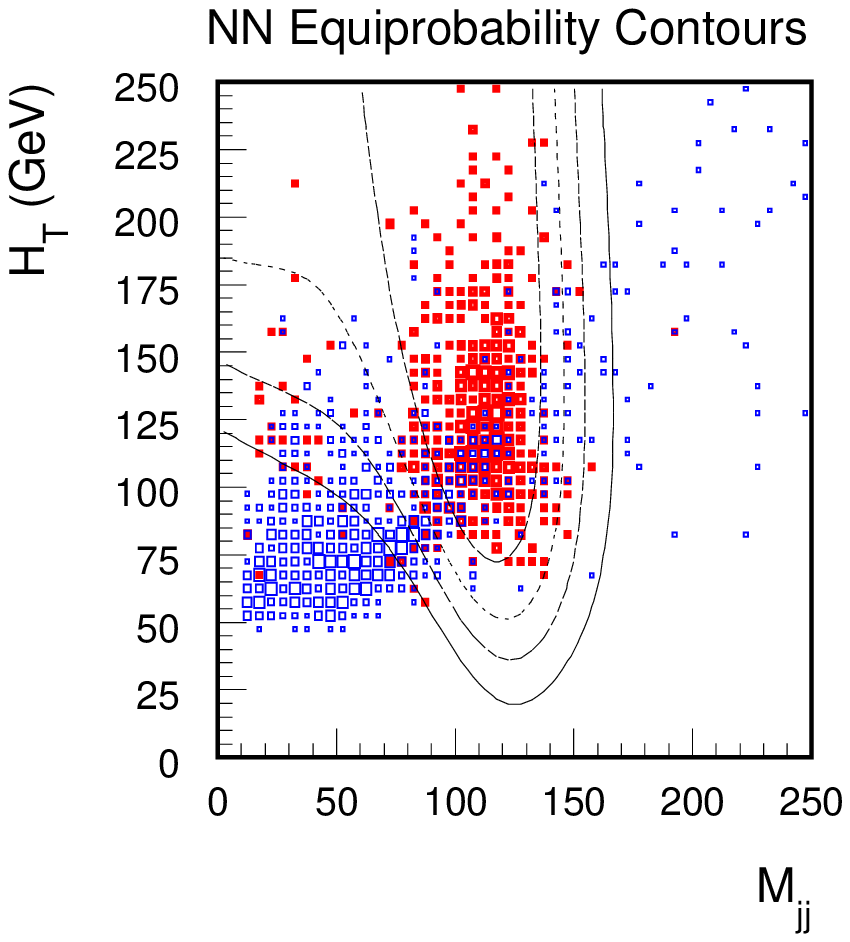}}
\caption[]{
\label{f-nn1}
\small The left panel shows predicted signal and background selection
 efficiencies from the neural network analysis.  Each point corresponds to a
 possible event selection.  The point labelled ``neural net'' is the result
 from the multivariate analysis described in the text.  The right panel shows
 neutral network output
 equal probability contours in the $H_T$ vs. $M_{jj}$ plane.  $H_T$ is the
 scalar sum of all jet energies, and $M_{jj}$ is the invarient mass of the
 tagged dijet system.  The open boxes are background events, and the closed
 boxes are signal.}
\end{figure}

In addition to these analyses, a multivariate analysis using neural networks
has been peformed for the $\ppWH\rightarrow\ell\nu\bb$ channel.  This style of
analysis has been used with considerable success by D\O\ in the top
mass\cite{tmass} and all--hadronic top decay analyses.\cite{thadronic} The
basic principle is to exploit correlations within an event in an automatic
manner.  The left panel of Fig.~\ref{f-nn1} shows the number of predicted
signal and background events for $\ppWH\rightarrow
\ell\nu\bb$ analyses.  Each point in the figure represents one possible 
analysis.  The band labelled ``rgsearch'' corresponds to hypothetical analyses
performed using selections using the standard technique of sequential
requirements applied to event variables, with each requirement a single--valued
comparison such as $\met>20$~GeV.  The point labelled ``TeV 2000'' is the
result from a previous Fermilab study\cite{Tev2K}.  The point labelled ``neural
net'' is the result from the multivariate analysis.  One sees that for a fixed
background, the signal is increased by roughly 50\% using the neural network.
Similar gains are expected in all other channels in this mass range.

\subsection{Other Improvements}
Results have also been obtained for hypothetical improvements in mass
reconstruction and $b$--jet tagging efficiency.  The analyses were repeated
after artifically improving the reconstructed dijet mass resolution in steps up
to a 50\% better resolution.  The results in Table~\ref{t-lowmass} include an
improvement in mass resolution of 30\%.  This level of improvement is possible
when information such as charged track energy is used in the mass
reconstruction in assocation with the calorimeter--based jet energies currently
used. Such an improvement has already been realized in a preliminary run I CDF
analysis of the $\pp\rightarrow Z\rightarrow\bb$ channel.\cite{zbb} Improved
mass resolution offers considerable benefits because for a selection with a
fixed signal expectation, the background will decrease as the resolution
improves.

The effect of improved $b$--jet tagging has also been explored by artificially
improving the second jet tagging efficiency by up to a factor of two.  The
gains from this improvement are not as important as those from mass resolution
improvements because both signal and background increase with improved tagging
efficiency.

\section{High Mass Higgs Searches, $M_H>135$~GeV}
Previous Fermilab studies have concentrated on the lower mass higgs states
which decay dominantly to $\bb$.  This study includes analyses designed
for final states in which the higgs decays to $WW$ or $ZZ$ instead of $\bb$.
This corresponds approximately to $M_H>135$~GeV.   Three final states are
considered:
(1) Three leptons, $\tl$, arising primarily from $\ppWH\rightarrow WWW$,
(2) Dileptons and neutrinos, $\dl$, from $\ppH\rightarrow WW$ and
(3) Like-sign dileptons plus jets, $\lsdl$, from $\ppWH\rightarrow WWW$
     and $\ppZH\rightarrow ZWW$.\cite{hmpub}
The dominant backgrounds are standard model production of $WW$, $WZ$, $ZZ$,
and $W(Z)+jets$ and $\ttb$ and multijet events with misidentification arising
from detector effects.  The standard model sources dominate the detector
effects.\footnote{In general, the backgrounds arising from detector effects
use conservative misidentification probabilities based on run I analyses by
both experiments.}

\begin{figure}	
   \centerline{\epsfysize=3in\epsfbox{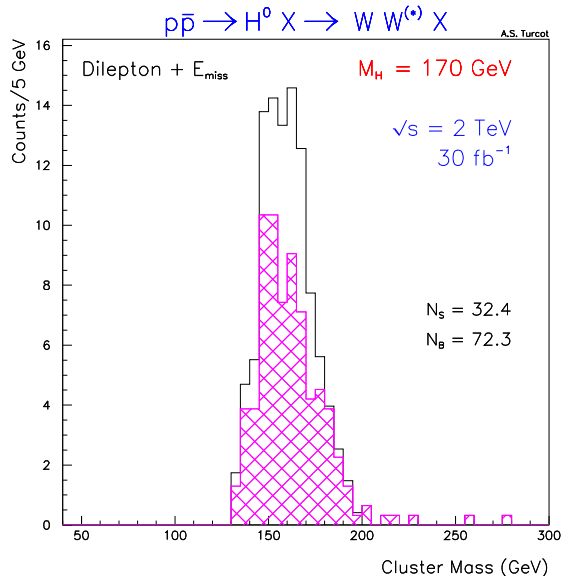}}
\caption[]{
\label{f-hm}
\small The cluster mass variable for background (shaded region) and for signal
 and background together (open region) for the $\dl$ analysis.}
\end{figure}

As for the low mass analyses, the initial selections are based on simple
variables related to the boson decay--product kinematics.  However, to reach
usable sensitivity, the analyses then use either (1) requirements typically
relating to angular correlations arising from spin differences between signal
and background or (b) likelihood methods.  In both cases new variables have
been designed.  Figure~\ref{f-hm} shows one such variable used in the $\dl$
analysis, the cluster mass $M_C\equiv\sqrt{p_T^2(\ell\ell) + m^2(\ell\ell)} +
|\met|$.  A result of the tuning is that the signal and background have similar
mass distributions, so these analyses must be treated as straight counting
experiments.  The numbers of expected signal and background events for the
high--mass channels are given in table~\ref{t-highmass}.

\section{Combination of Standard Model Search Channels}
\begin{figure}	
   \centerline{\epsfysize=4in\epsfbox{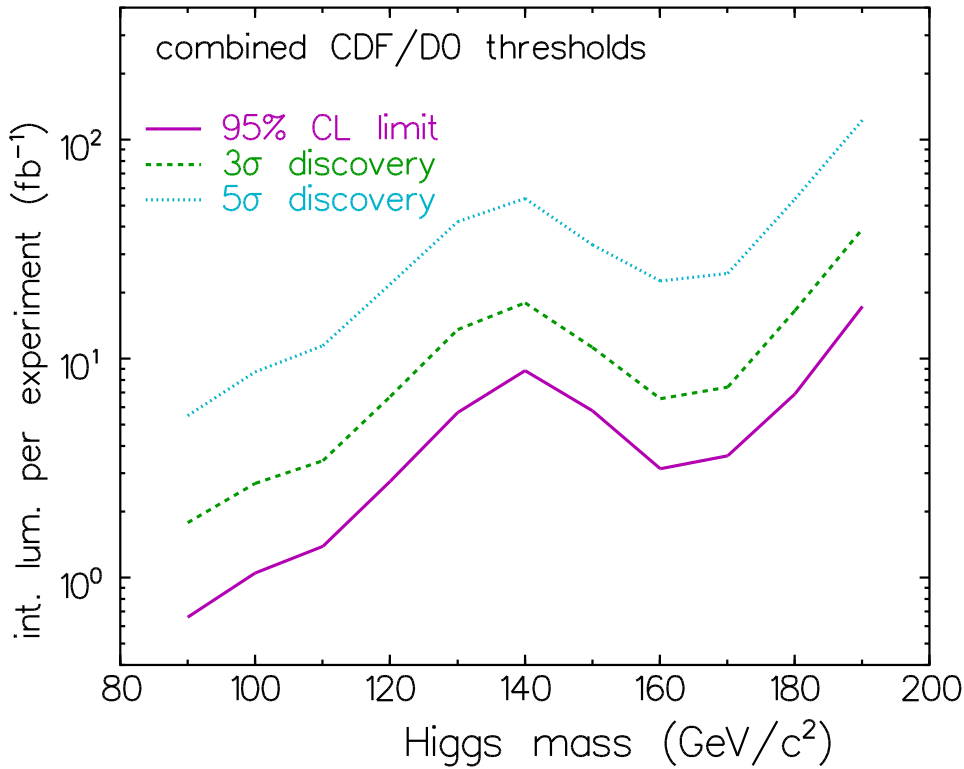}}
\vskip 2mm
\caption[]{
\label{f-combo}
\small Luminosity required to achieve 95\% confidence level exclusion,
3$\sigma$ evidence and 5$\sigma$ discovery as a function of higgs mass.  The
results use all channels and assume results from both CDF and D\O\ having
equal sensitivity.  The experimental uncertainties used are described in the
text.}
\end{figure}

The results in the preceeding two sections have also been combined to form a
single unified result.  Figure~\ref{f-combo} shows the luminosities required
for 95\% CL exclusion, 3$\sigma$ evidence and 5$\sigma$ discovery as a function
of standard model higgs mass.  These contours include statistical and
systematic errors\footnote{The systematic errors are assumed to scale with
luminosity.  The scaling is expected to hold at least until 2\% relative
systematic errors are reached.  Systematic uncertainties at this level do not
limit the analyses.} and the channels are combined using the prescription of
reference~\cite{shwstat}.

\section{Search for Neutral Higgs Bosons in SUSY Models}
\begin{figure}[t]	
  \centerline{\hbox{\epsfysize=3in\epsfbox{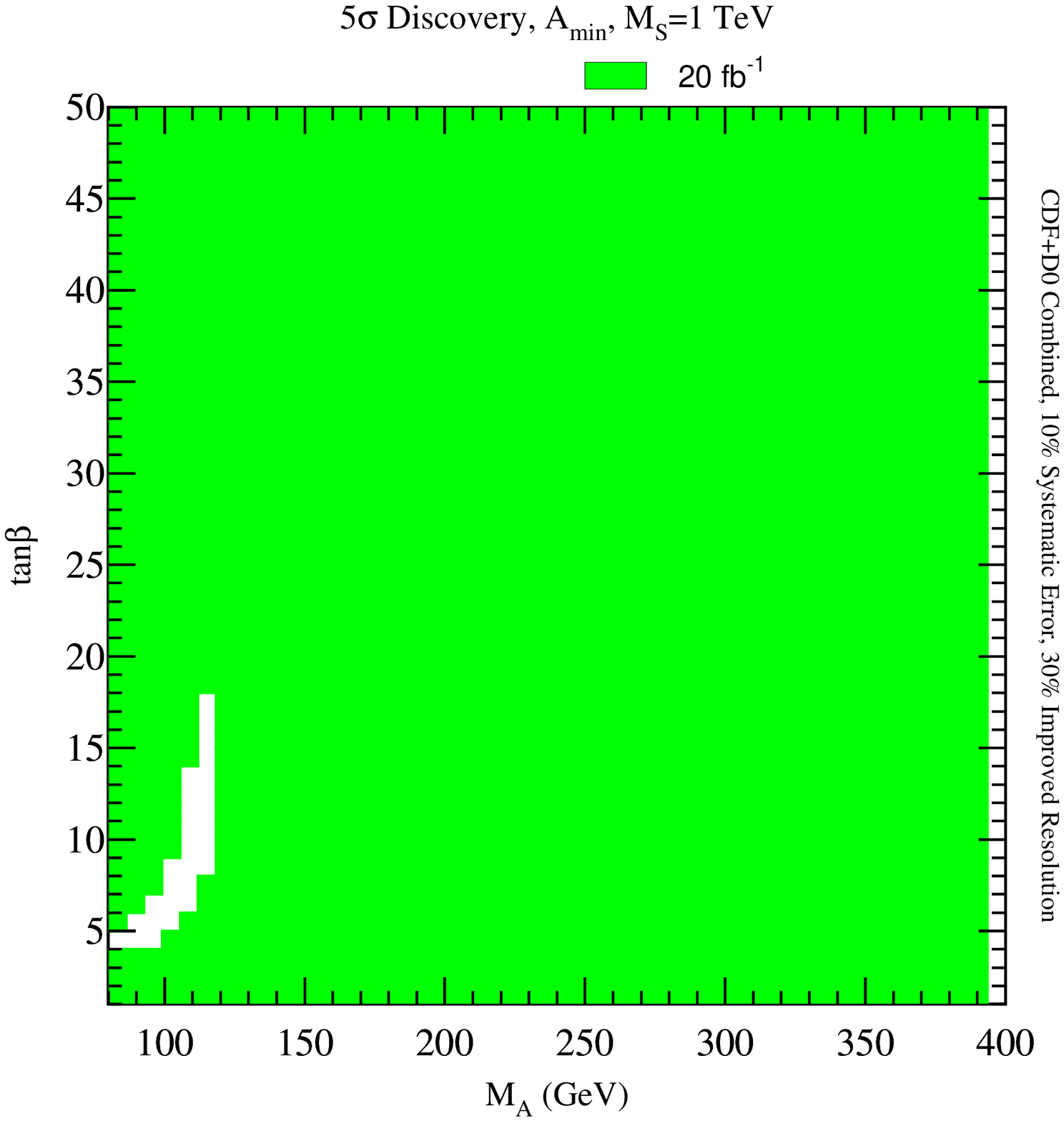}\hphantom{000}
    \epsfysize=2.6in\epsfbox{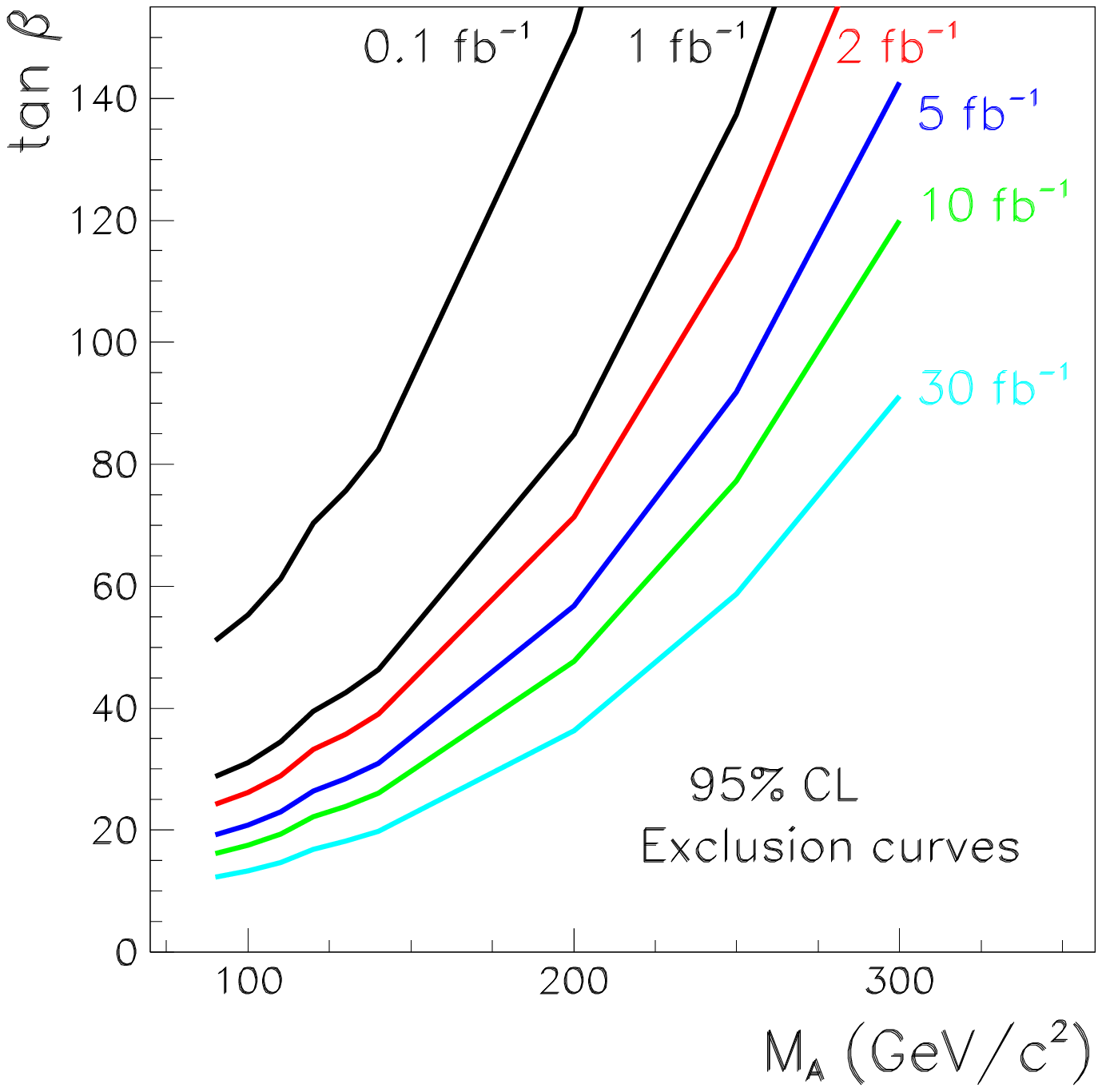}}}
\caption[]{\label{f-hsusy}
\small Exclusion contours for SUSY models.   For the left panel the shaded 
region is excluded based on a reinterpretation of the standard model searches
in section two as SUSY higgs production.  This example shows the 5$\sigma$
exclusion region with an integrated luminosity of 20~fb$^{-1}$.  The right
panel shows the result from the $\pp\rightarrow\phi\bb\rightarrow\bb\bb$
analysis described in section VII.}
\end{figure}

Most supersymmetric extensions of the standard model result in five physical
higgs states, two neutral scalars denoted $h$ and $H$(with $M_h<M_H$), a single
pseudoscalar $A$ and the charged doublet $H^\pm$.  Unlike most other
supersymmetric particles, the masses of the higgs particles depend only on two
parameters, typically chosen to be $\tan\beta$ and $M_A$ (or $M_h$).
Furthermore, the $h$ boson must satisfy $M_h<130$~GeV, and it may easily be
less massive that this.  For much of SUSY phase space, the $h$ decay is
identical to that of the standard model higgs, and only the production cross
section differs.  Given this, the searches for low mass standard model higgs
can be easily converted to searches for SUSY higgs.  The left--hand panel of
Fig.~\ref{f-hsusy} shows the 5$\sigma$ exclusion contour in the $\tan\beta$
vs. $M_A$ plane for the case in which all non-higgs SUSY masses are around 
1~TeV, the systematic error is 10\%, and CDF and D\O\ data are combined.
One sees that for this example, SUSY models having $1<\tan\beta<50$ and 
$80<M_A<400$~GeV are excluded.

\section{Searches for the SUSY Mode $\pp\rightarrow\phi\bb\rightarrow\bb\bb$,
  $\phi = h,\ A,\ H$} 
Within SUSY models, new higgs production modes exist, some with couplings
proportional to $\tan^2\beta$.  For sufficiently large values of $\tan\beta$
these become the dominant production modes.  One such mode is $\pp\rightarrow
h\bb$\footnote{There are additional modes with $A$ or $H$ in place of $h$.}.
This results in final states containing four $b$--flavored jets.  Analyses of
this channel have been carried out.  These typically require four jets, three
of which satisfy $b$--tag requirements.  All possible mass combinations of
$b$--jets are computed, and the resulting distribution is examined for a peak
near the generated higgs mass.  The 95\% C.L. exclusion contours in the
$\tan\beta$ vs. $M_A$ plane are shown in the right panel of Fig.~\ref{f-hsusy}
for a variety of integrated luminosities.

\section{Conclusions}
Studies of the experimental sensitivity to higgs production for Tevatron
Run II and beyond have been carried out.   Both standard model and
supersymmetric higgs production have been considered.   It is found that with
4~fb$^{-1}$ of data, standard model higgs can be excluded at 95\% confidence over
the interval $M_H<125$~GeV and $155<M_H<175$~GeV.   With 10~fb$^{-1}$, a 
standard model higgs boson will be seen as at least a $3\sigma$ excess over
the mass ranges $M_H<125$~GeV and $145<M_H<175$~GeV.  These results have been
converted to limits on the SUSY parameter space, with one example shown.

\begin{table*} 
  \caption[]{Numbers of expected signal and background events for each low--mass
   channel in 1~fb$^{-1}$.  A 30\% improvement in mass resolution over that from
   SHW has been assumed.  See the text for details.\label{t-lowmass}}
  \begin{tabular}{ccccccc} \tableline
   & & \multicolumn{5}{c}{Higgs Mass (GeV/c$^{-1}$)}\\
      Channel & Rate &  90 & 100 & 110 & 120 & 130 \\ \tableline
	      &  $S$ & 2.5 & 2.2 & 1.9 & 1.2 & 0.6 \\
   $\nn\bb$   &  $B$ & 10  & 9.3 & 8.0 & 6.5 & 4.8 \\
	      & \ssb & 0.8 & 0.7 & 0.7 & 0.5 & 0.3 \\ \tableline
	      &  $S$ & 8.4 & 6.6 & 5.0 & 3.7 & 2.2 \\
   $l\nu\bb$  &  $B$ & 48  &  52 & 48  & 49  & 42  \\
	      & \ssb & 1.2 & 0.9 & 0.7 & 0.5 & 0.3 \\ \tableline
	      &  $S$ & 1.0 & 0.9 & 0.8 & 0.5 & 0.3 \\
   $l^+l^-\bb$&  $B$ & 3.6 & 3.1 & 2.5 & 1.8 & 1.1 \\
	      & \ssb & 0.5 & 0.5 & 0.5 & 0.4 & 0.3 \\ \tableline
	      &  $S$ & 8.1 & 5.6 & 3.5 & 2.5 & 1.3 \\
   $\qq\bb$   &  $B$ & 6800& 3600& 2800& 2300& 2000\\
	      & \ssb & 0.10& 0.09& 0.07& 0.05& 0.03\\ \tableline
  \end{tabular}
 \end{table*}
 \begin{table*}
 \caption[]{Numbers of expected signal and background events in 1~fb$^{-1}$ for 
   the high--mass channels.  \label{t-highmass}}
\begin{tabular}{cc|ccccccc} \hline
 & & \multicolumn{7}{c|}{Higgs Mass (GeV/c$^{-1}$)} \\
 Channel & Rate &  120 & 130  & 140  & 150  & 160  & 170  & 180  \\ \hline
         &  $S$ & 0.04 & 0.08 & 0.11 & 0.12 & 0.15 & 0.10 & 0.09 \\
  $\tl$  &  $B$ & 0.73 & 0.73 & 0.73&  0.73 & 0.73 & 0.73 & 0.73 \\
         & \ssb & 0.05 & 0.09 & 0.13 & 0.14 & 0.18 & 0.12 & 0.11 \\ \hline
         &  $S$ &  --  &  --  & 2.6  & 2.8  & 1.5  & 1.1  & 1.0  \\
  $\dl$  &  $B$ &  --  &  --  &  44  & 30   & 4.4  & 2.4  & 3.8  \\
         & \ssb &      &      & 0.39 & 0.51 & 0.71 & 0.71 & 1.9  \\ \hline
         &  $S$ & 0.1  & 0.20 & 0.34 & 0.53 & 0.45 & 0.38 & 0.29 \\
 $\lsdl$ &  $B$ & 0.85 & 0.85 & 0.85 & 0.85 & 0.85 & 0.85 & 0.85 \\
         & \ssb & 0.11 & 0.22 & 0.37 & 0.57 & 0.49 & 0.41 & 0.31 \\ \hline
\end{tabular}
 \end{table*}

\end{document}